\documentclass[11pt]{article}
\usepackage{amssymb,cite,graphicx,a4}
\newcommand\blank[1]{}

\newcommand{\fract}[2]{{\textstyle\frac{#1}{#2}}}

\newcommand{\ri}{\right}

\newcommand{\lf}{\left}
\newcommand{\te}{\theta}

\newcommand{\balpha}{\alpha\kern -6.7pt\alpha}
\newcommand{\bbalpha}{\alpha\kern -4.95pt\alpha}

\newcommand{\CaC}{{\cal C}}

\newcommand\eq{\begin{equation}}
\newcommand\en{\end{equation}}
\newcommand\bea{\begin{eqnarray}}
\newcommand\eea{\end{eqnarray}}
\newcommand\nn{\nonumber}

\newcommand{\wt}{\widetilde}

\newcommand{\One}{{\hbox{{\rm 1{\hbox to 1.5pt{\hss\rm1}}}}}}
\renewcommand{\One}{{\mathbb 1}}
\renewcommand{\One}{{\rm 1\!\!1}}

\newcommand{\widetable}{\renewcommand{\arraystretch}{1.1}}
\newcommand{\ceff}{ c_{\rm eff}}
\newenvironment{tab}{\linespread{1.0} \begin{table}}{\end{table}%
\linespread{1.3}}
\begin{document}
\vskip 0.5cm
\begin{center}
{\Large \bf Massless flows between minimal $W$ models} 
\end{center}
\vskip 0.8cm
\centerline{Clare Dunning%
\footnote{\tt tcd1@york.ac.uk}
}
\centerline{\sl\small 
Dept.~of Mathematics, University of York, York YO1 5DD, UK }
\vskip 0.9cm
\begin{abstract}
\vskip0.15cm
\noindent
We study the renormalisation group flows between minimal $W$ models by
means of a new set of nonlinear integral equations which 
provide access to the effective central charge of both unitary and
nonunitary models.  We show that the scaling function associated to
the nonunitary models is a nonmonotonic function of the system size.  
\end{abstract}
\setcounter{footnote}{0}
\def\thefootnote{\fnsymbol{footnote}}

\noindent {\bf 1.}  A recent study of the renormalisation group
flows between  
nonunitary minimal models revealed an unexpected behaviour for 
 the groundstate energy $E_0(R)$, in
 that  it was a nonmonotonic  function of the system size
 $R$ \cite{DDT}. The nonmonotonicity was illustrated 
using the finite-size scaling function $\ceff(r)$, which  up to the
bulk term is proportional to the 
groundstate energy
\eq
E_0(R) = E_{\rm bulk}(M,R) - \frac{\pi \ceff(r)}{6 R} \quad, \quad  r=MR~,
\en
where $M$ is the so-called crossover scale (the mass in massive
theories). As 
the system size goes to zero  $\ceff(r)$ 
 becomes the effective central charge 
\eq
\lim_{r \to 0} \ceff(r) = c - 24 \Delta_0~.
\en
We denote the actual central charge by $c$ while  
$\Delta_0$ is the  
conformal dimension of the lowest primary field of
the UV CFT. 

The effective 
central charge and the central charge of the {\it unitary} minimal models
coincide, and according to 
Zamolodchikov's $c$-theorem \cite{Zamcth} there exists
a function $\wt c$ which is monotonic. However, apart
from the UV and IR points at which $\wt c$  equals the
central charge of the relevant  CFT, it is not clear if there is any
connection with 
$E_0(r)$. Nevertheless the groundstate energy of the unitary  models is always
monotonic. Analogously it had
been thought that the groundstate energy of the nonunitary models
would also be monotonic, but the 
results of \cite{DDT} and \cite{BH,FS,FSZ,Zd} 
provide a  number of counter examples. 
In this letter we study 
a further set of
perturbed  conformal field theories, demonstrating that 
 $\ceff(r)$ behaves nonmonotonically for the majority of nonunitary models.

We consider the  minimal models $W{\cal G}_N^{p,q}$ 
based on one of the   simply laced  Lie algebras
${\cal G}=A_{n-1},D_n,E_6,E_7,E_8$ \cite{BG}. The models are specified by two
coprime integers $p$ and $q$ with $p > h$, in terms of which the
central charge and the effective central charge are  
\eq
c=N\lf(1-\frac{h(h+1)(p-q)^2}{pq}\ri)\quad, \quad
\ceff=N\lf(1-\frac{h(h{+}1)}{pq}\ri)~. 
\label{ceff}
\en
Here $N$ denotes the rank of the algebra and $h$ the dual Coxeter
number. The primary fields $\Phi_{\Omega, \Omega'}$ are 
 labelled by  a pair of weights  $\Omega,\Omega'$ which satisfy
\eq
\te \cdot \Omega \leq q \quad,\quad \te \cdot\Omega' \leq p~,
\en
where $\te$ is the highest root of ${\cal G}$ \cite{LF}.
All models  have a 
primary field  $\Phi_{\rm adj}$ that has weights
$\Omega'=\Omega_{\rm id}$ and $  \Omega = \Omega_{\rm adj}$
corresponding to the 
trivial and adjoint representations of ${\cal G}$ respectively. It has 
conformal dimensions 
\eq
\Delta=\bar {\Delta}= 1-\frac{(q{-}p) h}{q},
\label{confdim}
\en
and is relevant for all $p,q$ such that $q>p.$
Formally denoting the action of the unperturbed CFT ${\cal A}_{\rm
  CFT}$, that
of the perturbed model may be written 
\eq {\cal A} = {\cal A}_{\rm CFT} +
\lambda \int \!\!d^2x \, \Phi_{\rm adj}~,  
\label{uvact}
\en 
 reproducing for $A_1$ the well-known $\phi_{13}$ perturbations of the
Virasoro minimal models. 

Depending on the sign of the
coupling constant $\lambda$, the perturbation 
either leads to  a massive
quantum field theory, or it induces a `massless flow'
into a 
conformal field theory with smaller effective central charge. 
One of the standard methods of studying the groundstate energy  of 
both types of 
model is the 
thermodynamic Bethe ansatz. The result is a set
of coupled nonlinear integral equations (the TBA equations), whose
solution provides direct access to $\ceff(r)$ at all values of
$r$. 

The  unitary $W{\cal
G}_N^{p,p+1}$  can alternatively be described as  
$ {\cal G}_N^k \times {\cal G}_N^1 /{\cal G}_N^{k+1}$ coset models at 
$k=p{-}h,$ in terms of which the  perturbing operator $\Phi_{\rm adj}$ is  
usually known as $\Phi_{\rm id,id,adj}.$
TBA equations describing the evolution of the effective central charge
between these coset models, and therefore the {\it unitary} $W{\cal
  G}$ models, are already known \cite{Zab,Maa,Ra1}, and they  
verify the conjectured 
pattern of flows  \cite{LF,CSS}: 
\eq
W{\cal G}_{N}^{p,p+1} +\Phi_{\rm adj} \to W{\cal G}_{N}^{p-1,p}~.
\label{wun}
\en 
By dropping the restriction $q
= p+1$, we may also consider flows originating from the much larger class
 of nonunitary $W{\cal G}$ minimal models, which also have a
 description as a coset but at fractional level ${\cal
   G}_N^{p/(q{-}p){-}h} \times 
   {\cal G}_N^{1}/ {\cal G}_N^{p/(q{-}p){-}h{+}1}.$  
Analogous to the known behaviour of
 the $\phi_{13}$ perturbations of the 
Virasoro models \cite{Za,LC,La,Aa}, it is natural to suppose that the
nonunitary 
generalisation of  (\ref{wun}) will be 
\eq 
W{\cal G}_{N}^{p,q} +\Phi_{\rm adj} \to W{\cal G}_{N}^{2p-q,p}~.
\label{wnonun}
\en

TBA equations describing massless flows from these nonunitary models are not
yet known. Instead,  
motivated by \cite{Zd,DDT}, we propose a 
 different type of nonlinear integral equation (NLIE) 
whose solution  
provides access to the effective central charge of 
both unitary and nonunitary minimal models. The equations can be found
in section 
 2, and are tested in section 3. In 4 and 5 we extract some exact
 results and make a comparison with 
ultraviolet and infrared perturbation  theory. The connection between the
massive models and  the Gross-Neveu models and other comments can be
found in section 6.

\vspace{0.2cm}
\noindent {\bf 2.} Our starting point is a set of nonlinear
integral equations that encode the 
groundstate energy of the 
imaginary-coupled simply laced affine Toda field theories. 
Our interest in these theories lies
in the fact that for  values of  the Toda coupling constant
$\beta^2=p/(p+1)$  
the theory can be consistently restricted 
to the massive $\Phi_{\rm adg}$
perturbation of the unitary $W{\cal G}_N^{p,p+1}$ minimal models
\cite{SmsG,lecl,RS,EY,BL,VF}.  
Moreover, the massive  $\phi_{13}$ perturbation of the nonunitary
$WA_1^{p,q}$ minimal  models can be obtained from the sine-Gordon
model by tuning the coupling  to  $\beta^2 =p/q.$
A similar result may be true for the 
 nonunitary minimal models 
based on the other simply laced Lie algebras \cite{KNS}, and at the level  
of the NLIEs we do find the choice
 $\beta^2 =p/q$   yields both  unitary and nonunitary
 perturbed models. 

The NLIEs describing the groundstate energy of the massive imaginary-coupled
Toda field theories were 
first obtained in \cite{Mar,Zi}, and have appeared in a
different context in 
\cite{DDT2}.   
The effective central charge is defined in terms of 
 $N$  functions which  satisfy a set of coupled equations
\eq
f^{(a)}(\te)=-\fract{i }{2} m_a r    e^{\te} +i \pi \sum_{b=1}
C^{-1}_{ab}\alpha_b +2i\sum_{b=1}^N \Bigl [\int_{\cal C} \!\!d\te' \,
\varphi_{ab}(\te-\te') \Im  m\ln (1+e^{f^{(b)}(\te')}) \Bigr ]~.
\label{fm}
\en
The integration contour ${\cal C}$ runs just below the real axis
while $r$
is built from the 
lightest mass $M$ of the theory and the cylinder size $R$ via
$r=MR$. We have set $m_a= M_a /M$, where each mass $M_a$  is associated
to a node of the Dynkin diagram via the  labelling of \cite{BFDS}, and
is such that 
$(M_1,M_2,\dots,M_N)$ forms an eigenvector of the Cartan matrix with
eigenvalue $4 \sin^2(\pi/2h)$.
Our  particular normalisation is given in table \ref{tab:masses}.  
\begin{table}
  \centering
\widetable
  \begin{tabular}{l|l}
\widetable
$ A_{n{-}1}$ & $  M_a =M \sin(\pi a/h)/\sin(\pi /h)$ \quad $1 \leq 
a\leq n-1$\\
$D_n$ &  $M_{n-1}=M_{n}=M/2\sin(\pi /h)  \quad, \quad M_a=M\sin(\pi a
/h)/\sin(\pi /h)   \quad 1 \leq a \leq n-2$ \\  
$E_6$  & $ M_1 = M_2 = M  \ , \ M_3 = 2M \cos(\pi /4)
\quad,\quad M_4=M_5 = 2M \cos (\pi/12) $ \\ 
&  $M_6 = 4M\cos(\pi/12) \cos(\pi/4)$ \\
$E_7$ & $M_1 = M \ , \ M_2 = 2M \cos(5\pi/18) \quad,\quad
M_3=2M\cos(\pi/9) \ , \  M_4 = 2M\cos(\pi/18)$  \\
& $ M_5 = 2M_2 \cos(\pi/18) \ , \ M_6 = 2M_3 \cos(2\pi/9) \ , \
M_7=2M_3 \cos(\pi/18)  $ \\
$E_8$ & $ M_1=M \ , \ M_2=2M \cos(\pi/5) \ , \ M_3 = 2 M \cos (\pi/30)
\ , \ M_4 = 2M_2 \cos(7\pi/30) $\\
& $ M_5 = 2M_2 \cos (2\pi/15) \ , \ M_6 = 2 M_2 \cos(\pi/30) $\\
& $ M_7 = 4 M_2 \cos(\pi/5) \cos(7\pi/30) \ , \ M_8 = 4 M_2 \cos(\pi/5)
\cos (2 \pi/15)$
\end{tabular}
  \caption{The mass normalisation.}
  \label{tab:masses}
\end{table}
The kernel functions 
\eq
\varphi_{ab}(\te)= \int_{-\infty}^{\infty} \frac{dk}{2\pi} \,e^{i k \te}
\lf(\delta_{ab} - \frac{ \sinh(\frac{\pi }{h}(\xi{+}1)k)) }{\sinh(\frac{\pi 
}{h}\xi k ) \cosh(\frac{\pi }{h}k )}C_{ab}^{-1}(k) \ri)
\en
are written in terms of  the `deformed' Cartan matrix $C_{ab}(k)$,
which is equal 
to $2$ if $a=b$, and $-1/\cosh(\pi k/h)$ if nodes $a$ and $b$
of the relevant Dynkin diagram are connected. Note that at $k=0$ it
reduces to  the standard Cartan matrix $C_{ab}$.  
The exact effective central charge may be determined using
\eq
c_{\rm eff}(r)= -\frac{6r}{\pi^2}\sum_{a=1}^{N} m_a \lf [
\int_{\CaC} \!\!d\te \sinh \te \, \Im m\ln (1+e^{f^{(a)}(\te)}) 
\ri ]~.
\label{cm}
\en

Due to the nonlinear nature of 
(\ref{fm}) and (\ref{cm}), $\ceff(r)$ is usually obtained by solving
the equations numerically.  
However, like the TBA equations, the NLIEs can be exactly evaluated at
the ultraviolet point \cite{DVuni}, the result for the above equations
being \cite{Zi}   
\eq
\ceff(0) = N-\frac{3 \xi}{\xi{+}1}\sum_{a,b=1}^N C_{ab}^{-1} \alpha_a
\alpha_b~. 
\label{ceff0}
\en
The effective central charge of the 
massive $\Phi_{\rm adj}$ perturbation of  $W{\cal G}_N^{p,q}$ is
obtained by setting 
the parameter $\xi$
and the twists ${\bf \alpha}=(\alpha_1 , \alpha_2,\dots, \alpha_N)$ to
\eq
\xi=p/(q{-}p) \quad ,\quad {\bf \alpha}=(2/p,2/p,\dots,2/p)~.
\label{xial}
\en
The affine Toda coupling constant $\beta^2$ is related to $\xi$ via
$\beta^2=\xi/(\xi+1)$, and the above ensures
$\beta^2=p/q.$  The choice of $\alpha$ is 
 motivated by \cite{FMQR,FRT3} for the $A_1$ related models, and the
 prescription given in  \cite{Zi} which yields
 the central charge  of the $W{\cal G}$ minimal models rather than $\ceff(r)$. 
As a first check we insert (\ref{xial}) into (\ref{ceff0}),
simplify using $12 \sum_{a,b=1}^N C_{ab}^{-1} =  
N h (h{+}1)$, and recover the expected UV effective central
charge 
(\ref{ceff}).

In accordance with (\ref{wnonun}) we modify the 
 massive equations to  interpolate  from  
a model  with $\xi = p/(q{-}p)$ to one with 
 $\xi'= (2p{-}q)/(p{-}q),$ that is $ \xi'=\xi{-}1$.  
Motivated by  \cite{Zd,DDT}, we associate two functions
$f_R^{(a)}$ and $f_L^{(a)}$  to each
node of the Dynkin diagram, introduce    new kernels and twists such
that the functions satisfy 
\bea
f_R^{(m)}(\te)&=&-\fract{i  }{2} m_a r   e^{\te} +i \pi \sum_{b=1}^N
C_{ab}^{-1}\alpha'_b  \nn\\ 
&&  \hspace{-2.5cm} +2i\sum_{b=1}^{N} 
\Biggl [ \int_{\CaC}\!\!d\te'\,\phi_{ab}(\te{-}\te')\Im m\ln
(1+e^{f_R^{(b)}(\te')}) 
+\! \!\! \int_{\CaC}\!\!d\te'\,\chi_{ab}(\te{-}\te')\Im m\ln
(1+e^{-f_L^{(b)}(\te')}) 
\Biggr ],  \\ 
f_L^{(m)}(\te)&=&-\fract{i  }{2} m_a r  e^{-\te} -i \pi
\sum_{b=1}^NC_{ab}^{-1}\alpha'_b  \nn\\ 
&&  \hspace{-2.5cm} -2i\sum_{b=1}^{N} \Biggl [  
\int_{\CaC}\!\!d\te'\, \phi_{ab}(\te{-}\te')\Im m\ln
(1+e^{-f_L^{(b)}(\te')}) +\!\!\!\int_{\CaC}\!\!d\te'\,
\chi_{ab}(\te{-}\te')\Im m\ln 
(1+e^{f_R^{(b)}(\te')}) \Biggr ].
\label{mL}
\eea
and replace the  formula for the effective central charge with
\bea
c_{\rm eff}(r)= -\frac{6r}{\pi^2} \sum_{a=1}^{N}  m_a \Bigl [
\int_{\CaC} \!\!d\te\, e^{\te} \Im m\ln (1+e^{f_R^{(a)}(\te)}) 
- \int_{\CaC} \!\!d\te\, e^{-\te} \Im m\ln (1+e^{-f_L^{(a)}(\te)})
\Bigl ]~.
\label{ceffnew}
\eea

As explained in \cite{DDT}, we fix the kernel functions by considering the
equations in the limits in which $r \to 0 $ and $r \to
\infty$. In the far infrared, the massless equations coincide with the
ultraviolet limit of the 
massive equations, with kernel $\varphi_{ab}(\te)$ replaced by
$\phi_{ab}(\te).$  Since this should describe  a model with
parameter $\xi{-}1$ 
we set $\phi_{ab}(\te)$ to
\eq
\phi_{ab}(\te)= \int_{-\infty}^{\infty} \frac{dk}{2\pi} \, e^{i k \te}
\lf(\delta_{mt} - \frac{ \sinh(\frac{\pi \xi}{h}k) }{\sinh(\frac{\pi
}{h}(\xi{-}1) k) \cosh(\frac{\pi}{h}k )}C_{mt}^{-1}(k) \ri)~.
\label{phi}
\en
For very small $r$, the massless equations 
should instead coincide with the ultraviolet limit of the massive
equations with parameter $\xi$.
After  some manipulations \cite{DDT}, the fourier 
transformed  
massive and massless equations can be directly compared. With the tilde
denoting  the
fourier transformed functions, the equations will match provided 
$\wt
\chi_{ab}(k)$ and $\alpha'_a$ satisfy
\bea
\wt \varphi_{ab}(k)&=&   \wt \phi_{ab}(k) + \sum_{c,d=1}^{N}
\wt \chi_{ac}(k)[\One- \wt\phi(k)]^{-1}_{cd}\, \wt  \chi_{db}(k)~,
\label{chiexp} \\
\alpha_a &=& \alpha_a' +\sum_{c,d=1}^N
\wt \chi_{ac}(0)[\One-\wt \phi(0)]^{-1}_{cd}   
\, \wt \chi_{db}(0)~. 
\eea
Inserting   the expressions for  $\wt
\varphi_{ab}$ and $\wt \phi_{ab}$ into (\ref{chiexp}) we  find  
\eq
\lf (\frac{ \sinh ( \frac{\pi }{h} k )}{\sinh( \frac{\pi
}{h}(\xi{-}1)k)\cosh(\frac{\pi }{h}k)}\ri)^2C_{ab}^{-1}(k) = \sum_{c,d=1}^N
\wt \chi_{ac}(k) \, C_{cd}(k) \, \wt \chi_{db}(k)~.
\en
After multiplying both sides by  $C_{ fa}(k)$ and summing over the index $a$,
we can 
write the right hand side as a square: $(\wt \chi(k) \, C(k) )_{fb}^2.$ Taking
the square root  and inverting the fourier transform
yields 
\eq
 \chi_{ab}(\te)=\pm \int \!\!\frac{dk}{2\pi}\, e^{i k \te} \frac{\sinh(\frac{\pi
   }{h}k)} {\sinh(\frac{\pi 
}{h}(\xi{-}1)k )\cosh(\frac{\pi }{h}k)}C_{ab}^{-1}(k)~.
\label{chi}
\en
The above only fixes $\chi_{ab}(\te)$ up to a
sign, but we find that choosing the negative sign results in an
effective central charge consistent with a  $W{\cal G}$ minimal
model, whereas the other 
choice does not yield a recognisable formula for $\ceff(0)$.  With the
negative sign we find  the new twists should be 
\eq
\alpha_a' = \frac{\xi }{\xi{-}1}\alpha_a~.
\label{alp}
\en
To avoid the pole in (\ref{alp}) at $\xi =1 $,  
and the poles in the kernels which 
cross the real axis as
$\xi$ falls below one, we  only consider models with  $2p
>q$ and therefore $\xi > 1$.
This is a sensible restriction 
since a flow of the form (\ref{wnonun}) with $2p < q$ would have an infrared 
CFT labelled by   $(2p-q,p)$, the first of which is negative.

\begin{figure}
$$
\begin{array}{cc}
\refstepcounter{figure}
\label{wa3}
\includegraphics[width=0.460\linewidth,height=0.46\linewidth]{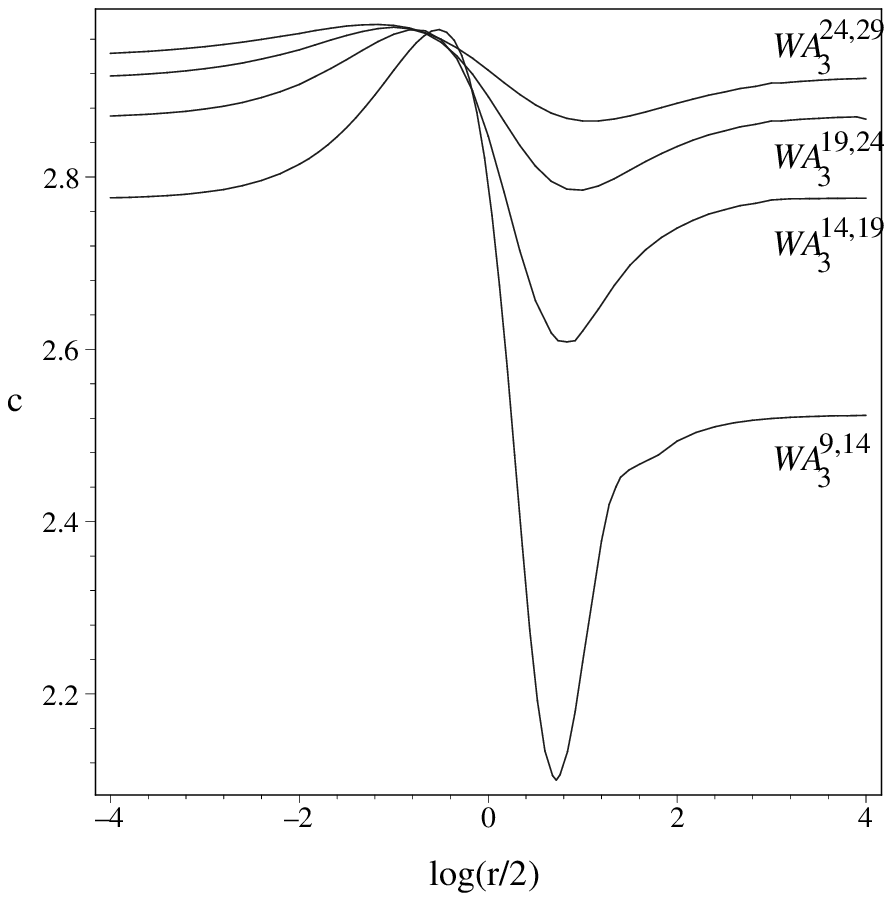}~~~~
&
\refstepcounter{figure}
\label{wd4}
\includegraphics[width=0.460\linewidth,height=0.46\linewidth]{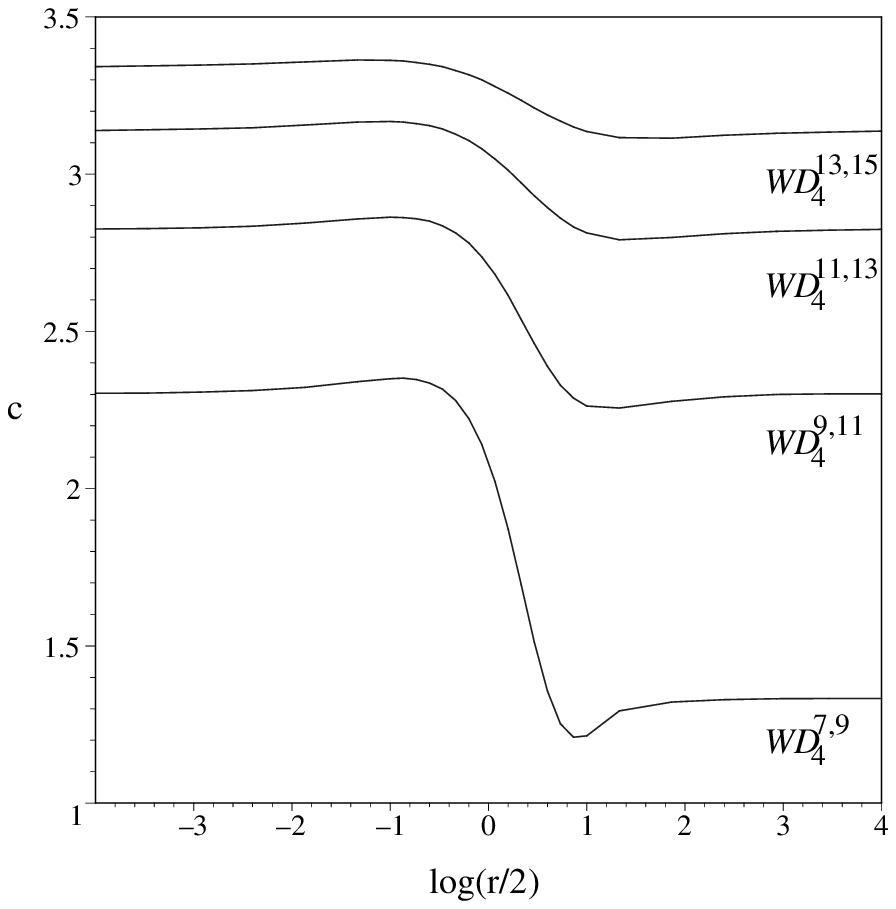}
\\
\parbox{0.460\linewidth}{
{\small Figure \ref{wa3}:
A family based on $WA_3$ with $J{=}6.$
}}
&
\parbox{0.460\linewidth}{
{ \small  Figure \ref{wd4}:
A family based on $WD_4$ with  $J{=}2$.
}}
\end{array}
$$
\end{figure}

\vspace{0.2cm}
\noindent {\bf 3.} The massless NLIEs have ultraviolet and infrared
values of $\ceff(r)$ which exactly match those of the conjectured flow
(\ref{wnonun}) provided we continue to use the 
massive prescription for $\xi$ and
$\alpha$ (\ref{xial}). We find 
\eq 
\ceff(0) = N\lf(1-\frac{h(h{+}1)}{pq}\ri)
\quad,\quad
\ceff(\infty) = N\lf(1-\frac{h(h{+}1)}{(2p{-}q)p}\ri)~.
\en
The massless flows naturally fall into families indexed by an integer 
$J=q{-}p:$ 
\eq
W{\cal G}_N^{p, p+J} +\Phi_{\rm adj} \to    W{\cal G}_N^{p-J, p}~. 
\en 
At $J=1$ there is a unique family corresponding to the flows
between the unitary minimal models. For these models we tested the
massless NLIEs against the  
the TBA equations \cite{Maa,Ra1}, typically finding very good
agreement. As expected the effective central charge was consistently
monotonic. For $J>1$ there are $\varphi(J)$ different families, each
of which interpolates between 
 nonunitary models (here 
$\varphi$ denotes the Euler-$\varphi$ function). Solving the NLIEs 
for the nonunitary models, we found that  $\ceff(r)$ 
 increases away from its UV value, undergoes a number of oscillations and
 then settles down to the predicted IR value. 
The  nonmonotonic 
behaviour of  two families of flows is illustrated in Figures 
\ref{wa3} and \ref{wd4}.

\vspace{0.2cm}
\noindent {\bf 4.} To strengthen the validity of the conjectured
equations we extract  a number of further 
predictions, which are then compared with  results from 
ultraviolet and infrared  conformal perturbation theory. 

The UV groundstate energy of a CFT perturbed by a
 primary field $\Phi$ of conformal dimension
$\Delta_{\rm UV}$  is predicted to behave
 as   \cite{ZamTBA,KMpest} 
\eq
E_0(R) = R B (\lambda)  - \frac{\pi c_{\rm pert}(r )}{6R} \quad,\quad
c_{\rm pert}(r)= \ceff(0)+\sum_{j=1}^\infty C_j \, (\lambda R^y)^j~,
\label{epert}
\en
where  $y=2(1-\Delta_{\rm UV})$ and 
the coefficients $C_j$ are proportional to the connected correlation
functions of the perturbing field on the plane.
 The action (\ref{uvact}) implies $\lambda$ and $M$ must satisfy
\eq
\lambda = \kappa M^y
\en
for a dimensionless constant $\kappa$, further implying that 
$c_{\rm pert}$ expands in powers of $r^y$.
On the other hand, the
 $2(\xi{+}1)i \pi /h$ periodicity \footnote{While the periodicity is easily
 extracted from the associated 
   Bethe ansatz equations \cite{Zi,DDT2}, it is less trivial to see
   directly from the NLIEs.} of the
nonlinear integral equations suggests that $\ceff(r)$ expands as a series in
$r^{2h/(1+\xi)}$, which will agree with  $c_{\rm pert}(r)$  provided  
 $2(1-\Delta_{\rm  UV}) = 2h/(1{+}\xi)$.  Substituting $p/(q{-}p)$ for
 $\xi $, we find the NLIEs predict  a 
 value of $\Delta_{\rm UV}$  which exactly matches 
the conformal dimension of $\Phi_{\rm adj}$ (\ref{confdim}).   

The expansion 
\eq
\ceff(r)=\ceff(0) + B(r) +\sum_{j=1}^\infty c_j \, (r^{y})^{j}~,
\label{cexp}
\en
differs from
  $c_{\rm pert}(r)$  by the bulk term $B(r)$, but  it 
  may easily be extracted 
from the NLIEs. For this we need 
 the leading asymptotics of the kernels as $\te \to
-\infty$.  The denominator of the inverse deformed
Cartan matrix  has either a term of the form $\sinh(\pi k)$ ($A_n$) or
$\cosh(\pi k/2)$ ($D_n,E_6,E_7,E_8$), both possibilities leading to a
simple pole in  
the kernels  
$\phi_{ab}(\te)$ and $\chi_{ab}(\te)$ at $k=-i$. 
Moreover, the kernels (\ref{phi}, \ref{chi}) have a  pole
at $k=-i h/(\xi{-}1)$, which contributes a term  proportional to 
$e^{h\te/(\xi{-}1)}$.  
Therefore we have 
\bea
\phi_{ab}(\te) \sim \phi_{ab}^{(1)} \, e^\te + p_{ab}^{(1)} \, e^{h
  \te/(\xi{-}1)} + \dots ,\quad \te \to -\infty ~, \label{asyp}\\
\chi_{ab}(\te) \sim \chi_{ab}^{(1)} \, e^\te + c_{ab}^{(1)} \, e^{h
  \te/(\xi{-}1)} + \dots ,\quad \te \to -\infty ~.\label{asyc} 
\eea
The  expansion coefficients can be 
  calculated  by taking the  appropriate residues, but
  we will only need  
  $\phi_{1a}^{(1)}$ and 
  $\chi_{1a}^{(1)}$, both of which can be written as
\eq
\phi_{1a}^{(1)} = \frac{\sin ( \frac{\pi }{h}\xi)\sin(\frac{\pi}{h})}{\pi
  \, \sin(\frac{\pi}{h
}(\xi{-}1)) \, \nu({\cal G})}  \, m_a~ \quad ,\quad
\chi_{1a}^{(1)} = \frac{\sin^2 ( \frac{\pi }{h})}{\pi
  \sin(\frac{\pi}{h}(\xi{-}1)) \,  \nu ({\cal  G})} \,m_a~, \label{phichi}
\en
where
$$
\nu(A_{n-1} )=1 \quad , \quad \nu(D_n) = 1/2 \quad,\quad 
\nu(E_6) =\sqrt {2/3} \sin (\pi/12)  
$$ 
\eq
\nu(E_7)= \frac{2}{\sqrt 3} \sin(\pi/18) \sin (2 \pi/ 9) 
\quad,\quad \nu(E_8)= \frac{ \sin (\pi/30)}{2\sqrt 3  \sin(\pi/5)}~.
\en
A small generalisation 
of the argument described in \cite{DDT} 
yields the massive and massless bulk terms:
\eq
B_{\rm massive}(r)=-\frac{3 \sin(\frac{\pi}{h}\xi) \, \nu({\cal G})
}{2\pi \sin(\frac{\pi}{h} 
  (1+\xi))\,  \sin (\frac{\pi}{h}) }  r^2 \quad , \quad
B_{\rm massless}(r)=\frac{3 \, \nu({\cal G})}{2\pi \sin(\frac{\pi}{h}
  (1+\xi))) }   r^2 ~.
\label{bulkM}
\en
The  bulk term corresponding to the massive perturbation of the unitary
minimal models has been calculated by 
Fateev \cite{Fatex} in the context of the associated coset. By
analytically continuing the coset parameter $k$ to rational values
(set $k=\xi{-}h$), we find Fateev's massive bulk term exactly
coincides with ours for both unitary and nonunitary models. 

The massless bulk terms are new, but we can
make at least one concrete check before turning to numerics: the models 
with $p=h{+}1$ and $q=h{+}2$ correspond to 
the coset $ {\cal G}_N^1 \times {\cal G}_N^1 /{\cal G}_{N}^2$,  for which the
massive and massless perturbations are known to coincide
\cite{Zab,DR}, as fortunately do our bulk terms at $\xi=h{+}1$.

The bulk terms have a simple pole
whenever $\xi{+}1=mh$ for some integer $m$, which should cancel against
one of the terms in the infinite sum  so that the 
expansion (\ref{cexp}) continues to be regular \cite{HF,CM,DTT}. 
In all cases the result, 
found by evaluating 
\eq
\lim_{\xi = mh{-}1} \frac{B(r)}{r^2}  [ r^2 - r^{2m
  \frac{h}{(\xi+1)}}]~,
\en
is a logarithmic term:
\eq
B_{\rm massive}(r)|_{\xi = mh{-}1}  = \frac{ 3  \,
  \nu({\cal G})}{\pi^2 m} r^2  
\ln r \quad , \quad
B_{\rm massless}(r)|_{\xi = mh{-}1}  = \frac{ 3 (-1)^m 
  \,\nu({\cal G})}{\pi^2 m} r^2 
\ln r~. 
\en
 
Now we are in position to compare (\ref{epert}) with
(\ref{cexp}), apart from one remaining difficulty. The 
perturbative coefficients $C_j$ are usually hard to calculate, and
instead it is easier to estimate, from the massless and massive NLIEs
respectively, the expansion coefficients $c_j$ and $\wt c_j.$
The perturbative coefficients $C_j$
do not depend on $\lambda$, and if we assume 
the mass and  
crossover scales $M$ are equal we should find 
\eq
c_j = (-1)^j \, \wt c_j~.
\en
We include a small sample of our numerics in table \ref{tab:WA710}.
Such good agreement 
provides an excellent check on both the massive and
massless NLIEs, the associated bulk terms and the above assumption
on $M$.  
\begin{tab}
\begin{center}
\begin{tabular}{ c| l| l   }   
\multicolumn{3}{c}{  $WA_2^{7,10}{+}\Phi_{\rm adj}$ } \vspace{0.2cm}
\\
               $j$ &   $~\wt{c}_j$              &   $~c_j$\\
\hline  
0 &~1.657142856           &~1.657142856           \\
1 &-1.49526585 &~1.49526587    \\
2 &~0.01009599 &~0.01009595         \\
3 &-0.0015271&~0.0015278         \\
4 &~0.0000603 &~0.0000602         \\ 
\end{tabular}
\qquad \qquad 
\begin{tabular}{ c| l| l   } 
\multicolumn{3}{c}{ $WD_4^{15,17}{+}\Phi_{\rm adj}$ } \vspace{0.2cm}   \\   
     
 $j$ &   $~\wt{c}_j$             &   $~c_j$     \\
\hline  
0 &~3.34117647           &~3.34117646           \\
1 &-0.1962515 &~0.1962514   \\
2 &~0.067572 & ~0.067573       \\
3 &~0.003602 & -0.003603      \\
4 &~0.00005 & ~0.000008      \\ 
\end{tabular}
\caption{
\protect{  Comparison of the massive and massless UV coefficients,
  found via the  NLIEs. } 
\label{tab:WA710}}
\end{center}
\end{tab}

\vspace{0.2cm}
\noindent {\bf 5.}  
Close to the infrared fixed point the 
model is described by the action of the 
infrared CFT plus an infinite number of contributions from  irrelevant
operators, resulting in a theory which is 
 unrenormalisable. However by
considering  the contribution of a finite number of fields it is
still 
possible to make a comparison with results from either NLIEs such as ours
or from TBA equations \cite{Zc,ZRSOS}.  We consider 
\eq
{\cal A} = {\cal A}_{\rm IR} +  g \int \psi \, d^2 x + t \int T {\bar
  T} \, d^2 x + \dots~,
\label{irexp}
\en
where the (possibly missing) irrelevant field $\psi$  of 
dimension $\Delta_{\rm 
  IR}$ and $T\bar T$ of dimension $2$ belong to the infrared CFT. The
couplings are related to the 
crossover scale $M$ as
\eq
g = \kappa_g M^{2-2\Delta_{IR}} \quad,\quad t = \kappa_t M^{-2}~.
\en
The action  implies $\ceff(r)$  has  IR expansion 
\eq
\ceff(r) \sim \ceff(\infty) + \sum_{j=1}^\infty g_j  (\kappa_g
r)^{(2-2\Delta_{\rm IR})j} +  \sum_{j=1}^\infty t_j  (\kappa_t
r)^{-2j}+\dots~.
\label{cinf}
\en

From the NLIE point of view \cite{DDT} corrections to $\ceff(r)$ come from the
$\te \to -\infty$ asymptotic of 
$\chi_{ab}(\te)$ given by (\ref{asyc}), the 
 $e^{h\te/(\xi{-}1)}$ term  generating a 
  a series of the form $r^{-2h/(\xi{-}1)}.$ Comparing with the CPT 
expansion (\ref{cinf}) leads to the prediction $\Delta_{\rm IR} =
1+h/(\xi{-}1)$, which  can be identified with the 
conformal dimension of the primary field $\Phi_{\rm   adj\,'}$ with
weights $\Omega=\Omega_{\rm id}, \ 
\Omega'=\Omega_{\rm adj}.$ 

We can also extract a  
prediction for $\kappa_t$ from the NLIEs.
 We start with 
the first two coefficients of the series generated by $T\bar T$
 \cite{ZRSOS}:
\eq
t_1 = -\frac{ \pi^3 \ceff(\infty)^2}{6} \quad,\quad
t_2 = \frac{\pi^6 \ceff(\infty)^3}{18}~, 
\label{ts}
\en
and compare them with the coefficients of $r^{-2}$ and $ r^{-4}$ found from
the NLIEs. Adapting the TBA argument  \cite{DTT}, we use the
$e^\te$ term in the 
expansion of $\chi_{ab}(\te)$ to  find
\eq
\ceff(r) \sim \ceff(\infty) - \frac{ 2 \pi^2 }{3} \ceff^2(\infty)
\chi_{11}^{(1)} r^{-2} + 2  \ceff^3(\infty) \lf(
\frac{2 \pi^2 }{3}  \chi_{11}^{(1)} \ri)^2r^{-4} +\dots~.
\label{cinfnl}
\en
The explicit form
\eq
\chi_{11}^{(1)}\,r^{-2}  = \frac{ \sin^2
  (\frac{\pi}{h})}{\pi \sin (\frac{\pi}{h} (\xi{-}1)) \nu ({\cal G})}r^{-2}
\label{chi11}
\en
indicates a pole whenever $\xi{-}1 =m'h $ for some integer
$m'.$  Evaluating as for the UV case we find
 (\ref{chi11})  becomes
\eq
\chi_{11}^{(1)}\,r^{-2} |_{\xi= m'h+1} =  -\frac{2 (-1)^{m'}
  \sin^2(\frac{\pi}{h})}{\pi^2 m' \nu({\cal G})}
  \,  r^{-2} \ln r~. 
\en
 The  infrared expansion coefficients 
  have been obtained  numerically for the models $WA_1^{p,p+1} + \phi_{13}$,  
$p=5,\dots,10$ in \cite{FQR,DTT}, and they show good agreement with
our predictions, while for  $WA_1^{p,p+1}$ and
$WD_{n+1}^{2n+2,2n+3}$ we find agreement with  the
theoretical results of \cite{DTT}. 
Finally, the effective central charge of the $\phi_{12},\phi_{21}$ and
$\phi_{15}$ perturbations of the Virasoro minimal models is exactly
half that of certain $WA_{2}$ models:
\bea
WA_1^{p,q} +\phi_{12} & \leftrightarrow & WA_2^{p,2q}+\Phi_{\rm adj}\nn \\
WA_1^{p,q} +\phi_{21} & \leftrightarrow & WA_2^{q,2p}+\Phi_{\rm adj} \\
WA_1^{p,q} +\phi_{15} & \leftrightarrow & WA_2^{2p,q}+\Phi_{\rm
  adj}~. \nn
\eea
Only the $\phi_{21}$ and $\phi_{15}$ perturbed models have  a 
massless flow, and the IR expansion coefficients  
found in \cite{DDT}  match  with those predicted above for the
associated $WA_2$ model. 
The correspondence actually works for any value of $\xi$ since the
$\phi_{12}/\phi_{21}/\phi_{15}$ NLIE 
is based on the tadpole diagram $T_1$, which is related by
folding to $A_2$.

Finally, comparing (\ref{cinfnl},\ \ref{chi11}) to (\ref{cinf},\
\ref{ts}) yields the promised prediction for $\kappa_t$:
\eq
\kappa_t = \frac{ 4 \sin^2 ( \frac{\pi}{h} )}{\pi^2
  \sin(\frac{\pi}{h}(\xi{-}1))\, \nu({\cal G})}~.
\en 

\vspace{0.2cm}
\noindent {\bf 6.} 
We have shown that the function $\ceff(r)$ for nonunitary models
interpolates from  a CFT in the ultraviolet to an infrared CFT that
has smaller  effective 
central charge, but in a nonmonotonic way.  It is likely that there is
a function, as yet unknown, which monotically interpolates 
between the nonunitary CFTs and satisfies a `nonunitary $c$-theorem'. 

We'd like to make two further comments concerning  the massive nonlinear
integral equations. First, level-rank (or KNS) duality
\cite{KNS,ABS} relates two 
nonunitary 
minimal models 
\eq
WA_{n{-}1}^{n,p} = WA_{p{-}n{-}1}^{p-n,p} \quad,\quad n,p {\rm \ coprime}~,
\en
which both have a field 
$\Phi_{\rm adj}$ with conformal dimension $\Delta=(pn{-}p{-}n^2) /p,$ 
resulting in entirely equivalent perturbed theories.
It is not at all obvious
that the NLIEs based on $A_{n-1}$ and those based on $A_{p{-}n{-}1}$ at the
appropriate values of $\xi$ will produce the same value of $\ceff(r),$ but
provided the normalisation of the lightest mass 
is such  that the coupling
$\lambda$ is the same for both models,    
our numerical studies confirm this. By equating the massive bulk
terms we find 
\eq
\sin( \pi/(p-n)) \   M_{[n{-}1]} = \sin(\pi/n) \ M_{[p{-}n{-}1]}~;
\en
the same result can also 
 be deduced  from  the relation  
$\lambda = \kappa M^y$ given for the 
{\it unitary} minimal models in \cite{Fatex}. 
We can rule out a 
massless flow from $WA_{n-1}^{n,p}$ via $\Phi_{\rm adj}$ by 
considering the conjectured IR model $WA_{n-1}^{2n-p,n}$, a sensible CFT
if   $2n{-}p>h $. Since 
 $h=n$ this would require  
 $p<n$,  contradicting the assumption $n<p$.  

As mentioned above, the massive $\phi_{15}$ and $\phi_{12}$
perturbations of the  
the Virasoro models 
 $WA_1^{2,p}$ (for odd $p$ greater than 4) are integrable, and the
 groundstate energy can be found using a 
 single NLIE based on $ A_2^{(2)}$ \cite{DT1}, with appropriate tuning of the
 free parameters.
Thus by  level-rank
 duality the models $WA_{p-3}^{p-2,p}$ have up to  two extra 
massive perturbations in addition to $\Phi_{\rm adj}$ whose
groundstate energy can be described in terms of a nonlinear integral equation. 

Second, the massive NLIEs encode the finite-size effects of a further set
of models obtained by sending $\xi=k{-}h $ to infinity.  In this limit
the perturbed coset  $ {\cal  
  G}_N^k \times {\cal G}_N^1 /{\cal   G}_N^{k+1} +\Phi_{\rm adj}$  
becomes the ${\cal G}$ Gross Neveu model  
\cite{ABL,VF}, with groundstate energy described in terms of an
infinite number ($k.N$)
of coupled TBA equations. The NLIEs offer a clear advantage over the
TBA  as the $N$ equations can 
still be solved numerically at any value of $r$. The kernels 
\eq
\varphi_{ab}(\te)= \int_{-\infty}^{\infty} \frac{dk}{2\pi} \,e^{i k \te}
\lf(\delta_{ab} - \frac{ e^{\pi |k|/h} }{\cosh(\frac{\pi }{h}k
    )}C_{ab}^{-1}(k) \ri) 
\en
form part of the prefactors of the associated S-matrices \cite{PF1}. 
The perturbation is (almost) marginal and the ultraviolet expansion
  of the effective central charge no longer has a simple power
  series form (\ref{cexp}). By studying the NLIEs (numerically and
  analytically) we hope to uncover the expected logarithmic
  corrections to $\ceff(r)$. If this approach is successful there are a  
 number of two-dimensional sigma models which 
have an interpretation as a perturbed  conformal field
theory \cite{PF1}. Since in all cases the  groundstate energy is
described by  
 an infinite number of  TBA equations, it 
would also be interesting to extend the current set of NLIEs to 
include such  models. 

Finally we remark that even though the groundstate energy of the
perturbed nonunitary models is apparently real, recent results
based on the $A_2^{(2)}$ models indicate the massive finite size
spectrum may 
in general 
be complex \cite{GW}.  
It remains an open question to study the excited states of  
the massless models, perhaps via the massless NLIEs, to see if a
similar result  holds.

\medskip
\noindent{\bf Acknowledgments --} 
I'm grateful to Pascal Baseilhac, Patrick Dorey,  Andr\'e LeClair,
Pierre Mathieu and  
Roberto Tateo for useful conversations.
I also thank the EPSRC for a Research Fellowship and IPAM, UCLA for
their hospitality.  

\end{document}